\documentclass[copyright]{eptcs}
\providecommand{\event}{PLACES 2017} 
\usepackage{breakurl}         
\usepackage{underscore}              

\usepackage{listings,xspace}
\usepackage{xcolor}
\usepackage{graphicx}
\usepackage{url}

\usepackage{bold-extra}

\lstdefinelanguage{Scala}%
{morekeywords={abstract,case,catch,char,class,%
    def,else,extends,final,%
    if,import,%
    match,module,new,null,object,override,package,private,protected,%
    public,return,super,this,throw,trait,try,type,val,var,with,implicit,%
    macro,sealed,%
  },%
  sensitive,%
  morecomment=[l]//,%
  morecomment=[s]{/*}{*/},%
  morestring=[b]",%
  morestring=[b]',%
  showstringspaces=false%
}[keywords,comments,strings]%

\newcommand{\comment}[1]{}

\newcommand{\ie}{{\em i.e.,~}}

\newcommand{\lacasa}{\textsc{LaCasa}}

\title{Towards an Empirical Study of Affine Types \\
for Isolated Actors in Scala}
\author{Philipp Haller
\institute{KTH Royal Institute of Technology\\ Stockholm, Sweden}
\email{phaller@kth.se}
\and
Fredrik Sommar
\institute{KTH Royal Institute of Technology\\ Stockholm, Sweden}
\email{fsommar@kth.se}
}
\def\titlerunning{Towards an Empirical Study of Affine Types for Isolated Actors in Scala}
\def\authorrunning{P. Haller \& F. Sommar}

\begin{document}
\maketitle

\begin{abstract}
LaCasa is a type system and programming model to enforce the object capability discipline in Scala, and to provide affine types. One important application of LaCasa's type system is software isolation of concurrent processes. Isolation is important for several reasons including security and data-race freedom. Moreover, LaCasa's affine references enable efficient, by-reference message passing while guaranteeing a ``deep-copy'' semantics. This deep-copy semantics enables programmers to seamlessly port concurrent programs running on a single machine to distributed programs running on large-scale clusters of machines. 

This paper presents an integration of LaCasa with actors in Scala, specifically, the Akka actor-based middleware, one of the most widely-used actor systems in industry. The goal of this integration is to statically ensure the isolation of Akka actors. Importantly, we present the results of an empirical study investigating the effort required to use LaCasa's type system in existing open-source Akka-based systems and applications.
\end{abstract}

\section{Introduction}

The desire for languages to catch more errors at compile time seems to
have increased in the last couple of years. Recent languages, like
Rust~\cite{Turon17}, show that a language does not have to sacrifice a
lot, if any, convenience to gain access to safer workable
environments. Entire classes of memory-related bugs can be eliminated,
statically, through the use of affine types. In the context of this
paper it is important that affine types can also enforce isolation of
concurrent processes.

\lacasa~\cite{HallerL16} shows that affine types do not necessarily
need to be constrained to new languages: it introduces affine types
for Scala, an existing, widely-used language. \lacasa~is implemented
as a compiler plugin for Scala 2.11.\footnote{See
  \url{https://github.com/phaller/lacasa}} However, so far it has been
unclear how big the effort is to apply \lacasa~in practice. This paper
is a first step to investigate this question empirically on
open-source Scala programs using the Akka actor framework~\cite{Akka}.

\paragraph{Contributions}
This paper presents an integration of \lacasa~and Akka. Thus, our
integration enforces isolation for an existing actor
library. Furthermore, we present the results of an empirical study
evaluating the effort to use isolation types in real applications. To
our knowledge it is the first empirical study evaluating isolation
types for actors in a widely-used language.

\paragraph{Selected Related Work}
Active ownership~\cite{ClarkeWOJ08} is a minimal variant of ownership
types providing race freedom for active objects while enabling
by-reference data transfer between active objects. The system is
realized as an extended subset of Java. Kilim~\cite{SrinivasanM08}
combines static analysis and type checking to provide isolated actors
in Java.  For neither of the two above systems, active ownership and
Kilim, have the authors reported any empirical results on the
syntactic overhead of the respective systems, unlike the present
paper. SOTER~\cite{NegaraKA11} is a static analysis tool which infers
if the content of a message is compatible with an ownership transfer
semantics. This approach is complementary to a type system which
enables developers to require ownership transfer
semantics. Pony~\cite{ClebschDBM15} and Rust~\cite{Turon17} are new
language designs with type systems to ensure data-race freedom in the
presence of zero-copy transfer between actors/concurrent processes. It
is unclear how to obtain empirical results on the syntactic overhead
of the type systems of Pony or Rust. In contrast, \lacasa~extends an
existing, widely-used language, enabling empirical studies.

\section{Background}\label{sec:background}

In this paper we study affine types as provided by
\lacasa~\cite{HallerL16}, an extension of the Scala programming
language. \lacasa~is implemented as a combination of a compiler plugin
for the Scala 2.11.x compiler and a small runtime
library. \lacasa~provides affine references which may be consumed at
most once. In \lacasa~an affine reference to a value of type \verb|T|
has type \verb|Box[T]|. The name of type constructor \verb|Box|
indicates that access to an affine reference is restricted. Accessing
the wrapped value of type \verb|T| requires the use of a special
\verb|open| construct:

\begin{lstlisting}[numbers=left, numberstyle=\scriptsize\color{gray}\ttfamily]
  val box: Box[T] = ...
  box open { x => /* use `x` */ }
\end{lstlisting}
\noindent
\verb|open| is implemented as a method which takes the closure
\verb|{ x => /* use `x` */ }| as an argument. The closure body then
provides access to the object \verb|x| wrapped by the \verb|box| of
type \verb|Box[T]|. However, \lacasa~restricts the environment (\ie
the captured variables) of the argument closure in order to ensure
affinity: mutable variables may not be captured. Without this
restriction it would be simple to duplicate the wrapped value,
violating affinity:

\begin{lstlisting}[numbers=left, numberstyle=\scriptsize\color{gray}\ttfamily]
  val box: Box[T] = ...
  var leaked: Option[T] = None
  box open { x =>
    leaked = Some(x) // illegal
  }
  val copy: T = leaked.get
\end{lstlisting}
\noindent
\lacasa~also protects against leaking wrapped values to global state:

\begin{lstlisting}[numbers=left, numberstyle=\scriptsize\color{gray}\ttfamily]
  object Global { var cnt: LeakyCounter = null }
  class LeakyCounter {
    var state: Int = 0
    def increment(): Unit = { state += 1 }
    def leak(): Unit = { Global.cnt = this }
    ...
  }
  val box: Box[LeakyCounter] = ... // illegal
  box open { cnt =>
    cnt.leak()
  }
  val copy: LeakyCounter = Global.cnt
\end{lstlisting}  
\noindent
The above \verb|LeakyCounter| class is illegal to be wrapped in a box.
The reason is that even without capturing a mutable variable within
\verb|open|, it is possible to create a copy of the counter, because
the \verb|leak| method leaks a reference to the counter to global
mutable state (the \verb|Global| singleton object). To prevent this
kind of affinity violation, \lacasa~restricts the creation of boxes of
type \verb|Box[A]| to types \verb|A| which conform to the object
capability discipline~\cite{Mill06a}. According to the object
capability discipline, a method \verb|m| may only use object
references that have been passed explicitly to \verb|m|, or
\verb|this|. Concretely, accessing \verb|Global| on line 5 is illegal,
since \verb|Global| was not passed explicitly to method \verb|leak|.

In previous work~\cite{HallerL16} we have formalized the object
capability discipline as a type system and we have shown that in
combination with \lacasa's type system, affinity of box-typed
references is ensured.

\begin{figure}
\begin{lstlisting}[numbers=left, numberstyle=\scriptsize\color{gray}\ttfamily]
  def m[T](b: Box[T])(implicit p: CanAccess { type C = b.C }): Unit = {
    b open { x => /* use `x` */ }
  }
\end{lstlisting}
\caption{Boxes and permissions in \lacasa.}
\label{fig:lacasa}
\end{figure}

Affine references, \ie references of type \verb|Box[T]|, may be
consumed, causing them to become unaccessible. Consumption is
expressed using {\em permissions} which control access to box-typed
references. Consuming an affine reference consumes its corresponding
permission.

Ensuring at-most-once consumption of affine references thus requires
each permission to be linked to a specific box, and this link must be
checked statically. In \lacasa~permissions are linked to boxes using
path-dependent types~\cite{AminGORS16}. For example,
Figure~\ref{fig:lacasa} shows a method \verb|m| which has two
formal parameters: a box \verb|b| and a permission \verb|p| (its
\verb|implicit| modifier may be ignored for now). The type
\verb|CanAccess| of permissions has a {\em type member} \verb|C| which
is used to establish a static link to box \verb|b| by requiring the
equality \verb|type C = b.C| to hold. The type \verb|b.C| is a
path-dependent type with the property that there is only a single
runtime object, namely \verb|b|, whose type member \verb|C| is equal
to type \verb|b.C|. In order to prevent forging permissions,
permissions are only created when creating boxes; it is impossible to
create permissions for existing boxes.

\begin{figure}
\begin{lstlisting}[numbers=left, numberstyle=\scriptsize\color{gray}\ttfamily]
class Box[T] { self =>
  type C
  def open(fun: T => Unit)
          (implicit p: CanAccess { type C = self.C }): Box[T] = {
    ...
  }
}
\end{lstlisting}
\caption{Type signature of the \texttt{open} method.}
\label{fig:signature-open}
\end{figure}

Since permissions may be consumed (as shown below), it is important
that opening a box requires its permission to be available.
Figure~\ref{fig:signature-open} shows how this is ensured using an
{\em implicit parameter}~\cite{OliveiraSCLY12} of the \verb|open|
method (line 5). Note that the shown type signature is simplified; the
actual signature uses a {\em spore} type~\cite{MillerHO14} instead of
a function type on line 4 to ensure that the types of captured
variables are immutable.

\paragraph{Consuming Permissions}
Permissions in \lacasa~are just Scala implicit values. This means
their availability is flow-insensitive. Therefore, changing the set of
available permissions requires changing scope. In \lacasa, calling a
permission-consuming method requires passing an explicit continuation
closure. The \lacasa~type checker enforces that the consumed
permission is then no longer available in the scope of this
continuation closure. Figure~\ref{fig:consuming-permissions} shows an
example.  \lacasa~enforces that such continuation-passing methods do
not return (see~\cite{HallerL16}), indicated by Scala's bottom type,
\verb|Nothing|.

\begin{figure}
\begin{lstlisting}[numbers=left, numberstyle=\scriptsize\color{gray}\ttfamily]
  def m[T](b: Box[T])(cont: () => Unit)(implicit p: CanAccess { type C = b.C }): Nothing = {
    b open { x => /* use `x` */ }
    consume(b) {
      // explicit continuation closure
      // permission `p` unavailable
      ...
      cont() // invoke outer continuation closure
    }
  }
\end{lstlisting}
\caption{Consuming permissions in \lacasa.}
\label{fig:consuming-permissions}
\end{figure}

\subsection{Akka}\label{sec:akka}

\begin{figure}
\begin{lstlisting}[numbers=left, numberstyle=\scriptsize\color{gray}\ttfamily]
class ExampleActor extends Actor {
  def receive = {
    case $msgpat_1$ =>
    $\ldots$
    case $msgpat_n$ =>
  }
}
\end{lstlisting}
\caption{Defining actor behavior in Akka.}
\label{fig:akka-actor}
\end{figure}

Akka~\cite{Akka} is an implementation of the actor
model~\cite{HewittBS73,agha86} for Scala. Actors are concurrent
processes communicating via asynchronous messages. Each actor buffers
received messages in a local ``mailbox'' -- a queue of incoming
messages. An Akka actor processes at most one incoming message at a
time. Figure~\ref{fig:akka-actor} shows the definition of an actor's
behavior in Akka. The behavior of each actor is defined by a subclass
of a predefined \verb|Actor| trait. The \verb|ExampleActor| subclass
implements the \verb|receive| method which is abstract in trait
\verb|Actor|. The \verb|receive| method returns a message handler
defined as a block of pattern-matching cases. This message handler is
used to process each message in the actor's mailbox. The \verb|Actor|
subclass is then used to create a new actor as follows:

\begin{lstlisting}[numbers=left, numberstyle=\scriptsize\color{gray}\ttfamily]
  val ref: ActorRef = system.actorOf(Props[ExampleActor], "example-actor")
  ref ! msg
\end{lstlisting}
\noindent
The result of creating a new actor is a reference object (\verb|ref|)
of type \verb|ActorRef|. An \verb|ActorRef| is a handle that can be
used to send asynchronous messages to the actor using the \verb|!|
operator (line 2).

\section{Integrating \lacasa~and Akka}\label{sec:overview}

\paragraph{The Adapter}
The LaCasa-Akka adapter\footnote{See \url{https://github.com/fsommar/lacasa/tree/akka}} is an extension on top of Akka. During its design, an important constraint was to keep it separate from Akka's internals -- primarily to limit the effect of internal changes as Akka evolves.

The adapter consists of two parts: \texttt{SafeActor[T]} and \texttt{SafeActorRef[T]}, both with the same responsibilities as their counterparts in the Akka API. However, note that in contrast to the latter, they are generic over the message type. Akka instead relies on pattern matching to discern the types of received messages (see Section~\ref{sec:akka}). For the LaCasa-Akka adapter, however, it is necessary to know the types of messages at compile time, to prevent the exchange of unsafe message types.

\begin{figure}
\begin{lstlisting}[numbers=left, numberstyle=\scriptsize\color{gray}\ttfamily]
trait SafeActor[T] extends Actor {
  def receive(msg: Box[T])(implicit acc: CanAccess { type C = msg.C }): Unit
}
\end{lstlisting}
\caption{Usage of \lacasa's boxes and permissions in \texttt{SafeActor}.}
\label{fig:safeactor}
\end{figure}

\paragraph{SafeActor}
A subclass of Akka's \texttt{Actor}, \texttt{SafeActor} provides a different \texttt{receive} method signature, which is the primary difference between the two. Instead of receiving an untyped message, of type \texttt{Any}, \texttt{SafeActor[T]} receives a boxed message of type \texttt{T}, and an access permission for the contents of the box (see Figure~\ref{fig:safeactor}).

\paragraph{SafeActorRef}
The API for \texttt{SafeActorRef} is a wrapper of Akka's \texttt{ActorRef}, and contains a subset of the latter's methods and functionality. It uses the same method names, but method signatures are different, to include necessary safety measures. For every method accepting a box, there is a dual method accepting a box and a continuation closure. Recall that it is the only way to enforce that boxes are consumed (see Section~\ref{sec:background}). The dual methods use the \texttt{AndThen} suffix to indicate that they accept a continuation closure.

For message types that are immutable, the API can be significantly simplified, resembling that of a regular Akka \texttt{ActorRef}. Meanwhile, internally, the message is still boxed up and forwarded for handling by the \texttt{SafeActor}. Importantly, though, the box does not have to be consumed, enabling the method to return and continue execution -- removing the need for the \texttt{AndThen} family of methods.

\section{Empirical Study}

We converted several Scala/Akka programs to use the LaCasa-Akka
adapter described in Section~\ref{sec:overview}. The goal of this
conversion is to evaluate the effort required to use \lacasa's type
system in practice. The converted programs are part of the Savina
actor benchmark suite~\cite{ImamS14a}. Concretely, we converted the
following programs: (1) In \verb|ThreadRing|, an integer token message
is passed around a ring of N connected actors. This benchmark is
adopted from Theron~\cite{TheronThreadRing}; (2) \verb|Chameneos| is a
micro-benchmark measuring the effects of contention on shared
resources while processing messages; (3) \verb|Banking| is a bank
transaction micro-benchmark measuring synchronous request-response
with interfering transactions.

\begin{table*}[t!]
  \begin{tabular}{ l | r | r | r | r }
  \hline
  {\bfseries ~Program~} & {\bfseries ~LOC (Scala/Akka)~} & {\bfseries ~LOC (\lacasa/Akka)~} & {\bfseries ~Changes~} & {\bfseries ~Changes (\%)~} \\
  \hline
  ThreadRing            & 130                            & 153                         & 27 add./10 del.                 & 28.5\%                     \\
  Chameneos             & 143                            & 165                         & 26 add./7 del.                  & 23.1\%                     \\
  Banking               & 118                            & 135                         & 27 add./12 del.                 & 33.1\%                     \\
  \hline
  \textbf{Average}      & 130                            & 151                         & ~                               & 28.2\%                     \\
  \hline
  \end{tabular}
  \caption{Results of the empirical study.}\label{tab:results}
\end{table*}

Table~\ref{tab:results} shows the results. On average 28.2\% of the
lines of code of each program needed to be changed (we exclude changes
to imports). It is important to note that we expect this number to be
significantly lower for larger applications where sequential,
non-actor-based code dominates the code base. The most important
reasons for code changes are (a) the declaration of safe message
classes and (b) the insertion and removal of messages into/from
boxes. For example, in \verb|ThreadRing| 33.3\% of added lines are due
to declaring message classes as safe.

\section{Conclusion}\label{sec:conclusion}

\lacasa~extends Scala's type system with affine types, with
applications to race-free concurrent programming and safe off-heap
memory management. This paper shows how \lacasa~can ensure the
isolation of actors in Akka, a widely-used actor framework for Scala,
while providing safe and efficient ownership transfer of asynchronous
messages. According to our empirical study, adjusting existing
Akka-based Scala programs requires changing 28.2\% of the lines of
code on average. However, this initial result represents a worst-case
scenario, since the study only considered micro-benchmarks where
actor-related code dominates, unlike larger real-world applications.
An empirical study extending our results to medium-to-large
open-source code bases is ongoing.

\bibliographystyle{eptcs}
\bibliography{bibliography}
\end{document}